# Lecturers' perspectives on the integration of research data management into teacher training programmes

Sandra Schulz[1], Juliane Jacob[2]

[1] University of Hamburg, Faculty of Education, Computer Science Education, University of Hamburg, Germany

[2] University of Hamburg, Center for sustainable Research Data Management, University of Hamburg, Germany

Corresponding author: sandra.schulz@uni-hamburg.de

## Introduction

Data is omnipresent nowadays and data literacy (DL) "emphasizes the ability to understand, use, and manage science data"(Qin & D'Ignazio, 2010). Outside of Germany, various programs and data literacy education (DLE) have been implemented much earlier and are now much more advanced already (Ridsdale et al., 2015). In Germany, the DL-Charter defines DL as follows: "Data literacy encompasses the skills to collect, manage, evaluate and apply data in a critical way"[1], whereby the aspects of research data management (RDM) data literacy in the sense of "research/scientific data literacy" are distinguished from "everyday data literacy" (Wuttke & Helling, 2020). This definition shows that research data management is a vital piece of the DL puzzle that students could benefit from integrating into their studies to develop or expand their skills. In practice, students work with empirical data and usually write Bachelor's and Master's theses without any prior knowledge of RDM. The University of Hamburg has had RDM projects for some time and the Centre for Sustainable Research Data Management (CRDM)[1] was established in 2017, which offers a wide range of RDM services. Training materials from the University of Hamburg and other institutions are also available online. Furthermore, the University of Hamburg has established a Data Literacy Education (DLE) network[2] since 2018, which in turn is a member of the Stifterverband[3] network of the same name. It sets the goals to enhance data literacy education and to offer learning opportunities for data literacy education. The demand for RDM training courses is increasing, especially from graduate schools (of the four University of Hamburg cluster of excellence[4]), and consultations often are used by applicants regarding the creation of data management plans (DMP). The RDM services such as the Research Data Repository (RDR)[5] are especially used by senior scientists to publish and archive research data or the Research Information System (RIS)[6] to provide evidence of publications and project funding.

---

[1] Center for sustainable research data management at University of Hamburg, last access 03.02.2025, https://www.fdm.uni-hamburg.de/en.html.

[2] Data Literacy Network University of Hamburg, only available in German, last access 03.02.2025, https://www.uni-hamburg.de/dle-netzwerk.html.

[3] Stifterverband, last access 03.02.2025, https://www.stifterverband.org/english.

[4] Four Clusters of Excellence at University of Hamburg from 2019, last access 03.02.2025, https://www.uni-hamburg.de/en/forschung/forschungsprofil/exzellenzcluster.html.

[5] Research Data Repository of University of Hamburg, last access 03.02.2025, https://www.fdr.uni-hamburg.de.

[6] Research Information System of University of Hamburg, last access 03.02.2025, https://www.fis.uni-hamburg.de/en.html.

Students often are not familiar with the concept of RDM, so they rarely search for RDM services and are rarely able to develop an RDM strategy in the sense of a shortened DMP. If they do find offers, the information and content is often too abstract and not applicable to the needs and prior knowledge of students, as it is primarily aimed at researchers from the doctoral stage onwards who have more expertise in dealing with data and the need to publish.

There is a discrepancy between the expectation that students should be "data literate" and the fact that usually, there is no well-organized/structured pathway within their studies to achieve this competency. Closing this gap was the motivation for the two projects "Early Education in Data Management Decisions" ($E^2D^2$) (in the Faculty of Education) and "Early Education in Data Management Decisions - an adapted course" ($E^2D^2$ adapted). As part of $E^2D^2$, students were specifically taught basic RDM knowledge to strengthen their DL skills. RDM potentially can improve the handling with data and the data quality, not only for scientists but also students. In particular, reflecting on the data life cycle and developing a DMP can be helpful in order to deal with legal aspects and consider archiving options at an early stage, for example. It also reduces students' dependency on their supervisors as they dispose of sufficient knowledge themselves. The status quo before the $E^2D^2$ projects was that there had been no demand for RDM from students/lecturers or that respective offers had not reached students or had been too abstract. Consequently, the $E^2D^2$ projects addressed these issues in a structured way. Another focus of the project was to conduct interviews with different lecturers.

## Theoretical Background

The MSDSE Institute (UC Berkeley Institute for Data Science, NYU Center for Data Science, UW eScience Institute) published a report (The Moore-Sloan Data Science Environments: New York University, UC Berkeley, And The University Of Washington, 2018) on the topic of data science, in which research data management is described as a core competence (alongside statistics and inference, machine learning, scalable/cloud computing, data visualisation, and ethics including security, privacy, and human contexts of practice) for students.

The necessity of RDM competencies have been "documented" and implemented at various universities/institutes (Carlson et al., 2015; Piorun et al., 2012), and their necessity is not only concentrated in areas that operate with big data, but also in the field of chemistry (Reisner et al., 2014), for example.

Carlson et al. (Carlson et al., 2013) designed a data information literacy project to identify the educational needs of graduate students, where they focused on a variety of science disciplines and responded with effective educational interventions. For this purpose, students and faculty members were interviewed to identify the specific requirements in data management education.

In addition to asking how advanced their knowledge (data generation and management) was, the interviewers also wanted to find out how important it was for graduate students to become knowledgeable. Faculty members and students were asked to rate the importance of twelve data information literacy competencies (data processing and analysis, data management and organization, data preservation, databases and data formats, ethics and attribution, data quality and documentation, data curation and reuse, data conversion and interoperability, data visualization and representation, discovery and acquisition, metadata and data description, cultures of practice) using a five-point Likert scale (1 = not important, 2 = somewhat important,

3 = important, 4 = very important, 5 = essential). Subsequently, they were asked to explain their choices. Interviewees were also allowed to name competences that could help them complete their studies.

Students and faculty members evaluated the average ranking of the twelve data information literacy competencies as "important" or higher (3.80 to 4.44, n = 25).

Other studies (e.g. American Library Association (ALA), Association for College and Research Libraries (ACRL), 2000; Calzada Prado & Marzal, 2013) summarize five competencies (like understand, find, evaluate, use and managing data). However, the results remain comparable: data literacy is a key skill for students during their studies.

However, a discrepancy is also described between the skills that students in graduate laboratories believe they should learn in their undergraduate education and the skills that these students possess.

Shorish (2015) described the evolution and background of the field of Data Information Literacy and changes in the scholarly landscape.

There are various programs that implement data literacy and data management in a structured way. One option are Data Carpentry workshops (Teal et al., 2015). However, this workshop model is tailored to the training needs of researchers. In order to also serve as a helpful resource for students, it would have to be adapted to their specific needs.

In German-speaking universities and comparable institutes, DLE and RDM activities have also increased. However, this primarily focusses on senior scientists (NFDI, state initiatives so-called Landesinitiativen), but increasingly, programmes are also being implemented for students [7][8][9][10][11]. (Wiljes & Cimiano, 2019) described RDM as a rather new topic in academic education. In 2013, they started to organize a seminar, at a time when there was no structured material (no textbook or teaching experiences) available. In 2019, there were some handbooks on RDM (Briney, 2015; Corti et al., 2014; Ludwig & Enke, 2013; Pryor, 2012; Ray, 2014), an Open Science book (Nielsen, 2014), online resources, online tutorials, Webinars and even a MOOC.

Even if more materials are available, teachers must adapt the content to the students' needs. And, which is also a challenge, the content has to fit into the curriculum in terms of time. Hunt (2005) describes the challenges of integrating Data Literacy into the curriculum in an undergraduate institution. Because lecturers are faced with the daily challenge of having to choose between the subject-specific knowledge they must impart, and the fundamental knowledge (e.g. "Good scientific Practice" or legal aspects) required for their students to complete their thesis. In this respect, the survey provided valuable insights into the general mood within the educational science department at the University of Hamburg.

---

[7] Forschungsdatenmanagement für Studierende, only available in German, last access 03.02.2025, https://www.ibi.hu-berlin.de/de/studium/rundumdasstudium/fdm-fuer-studierende.
[8] Certificate Course in Research Data Management receives the State Teaching Award 2023, last access 03.02.2025, https://www.fh-potsdam.de/en/news-media/news/certificate-course-research-data-management-receives-state-teaching-award-2023.
[9] Zertifikatskurs Forschungsdatenmanagement für Studierende, only available in German, last access 03.02.2025, https://www.uni-potsdam.de/de/veranstaltungen/detail/2024-02-26-zertifikatskurs-forschungsdatenmanagement-fuer-studierende.
[10] Forschungsdaten der Abschlussarbeit smart managen: Dein Weg zum Datenprofi, only available in German, last access 03.02.2025, https://www.ub.uni-leipzig.de/service/workshops-und-online-tutorials/forschungsdaten-der-abschlussarbeit-smart-managen-dein-weg-zum-datenprofi.
[11] Research data management for students, last access 03.02.2025, https://www.zml.kit.edu/english/7056.php

The RDM content was taught in accordance with the learning objective matrix (Petersen et al., 2023) in combination with the didactic approach of the proven Train-the-Trainer Concept on Research Data Management (Biernacka et al., 2020). This also corresponds to the course plans proposed by (Kafel et al., 2014) for an introduction to data management for undergraduate students. Based on these concepts, RDM learning objectives were taught in a structured way so that the students could expand their skills but were not "overwhelmed at the same time". In terms of content, students were introduced to the data life cycle and all including steps. Employing a practical approach that incorporated their own experiences, students worked on examples using their own data and projects. This directly contributed to their progress on homework and thesis projects. It also prepared them for the variety of ways of dealing with data and the fact that a lot is happening in the field. This in turn showed that it is worth learning new ways of handling data on your own and/or asking for support at contact points such as the Center for sustainable research data (CRDM) at the University of Hamburg.

Due to the lack of previous scientific work and few proven concepts for integrating RDM into the curriculum, we present here an exploratory study dedicated to the following research questions:

1. What role does RDM content play for lecturers in their current courses and supervision of students in teacher training?
2. What RDM content would lecturers like to expect from student in teacher training?
3. What RDM content would lecturers like to expect from themselves?
4. Which RDM teaching formats could be integrated into the curriculum?

## Methodology

A qualitative approach was chosen to answer the research questions to obtain detailed information about the relevance of RDM for lecturers and students. To this end, interview requests were sent to various lecturers at the Faculty of Education.

**Composition and justification of the cohort:** three university lecturers were interviewed who had already gained experience with RDM through third-party funded projects and in teaching. Of the three interviewees, one was female and two were male. All interviewees were lecturers at the Faculty of Education at the University of Hamburg at the time of the survey and taught students in various teacher training programmes or in the Educational Science degree programme.

The size of the cohort is sufficient since the aim of qualitative research is about showing different characteristics that are not weighted (instead of making generalized statements). Furthermore, Patton (2002) writes concerning the sample size in qualitative research: "There are purposeful strategies instead of methodological rules. There are inquiry approaches instead of statistical formulas." Dukes (1984) proposes that 3 to 10 interviewees are sufficient in qualitative research. In respect to our research goals to get first impressions on how to integrate RDM into students' curriculum and to better understand related challenges, we decided to conduct 3 interviews for this study.

**Interview procedure:** the interview was conducted via Zoom. Beforehand, the respondents signed an informed consent form, which informed them about the study and the data processing. The interviews were conducted by a researcher of the project. Its duration varied between 16 to 26 minutes.

**Data processing procedure:** The interviews were recorded and transcribed and then analysed using qualitative content analysis (Mayring, 2010) and the MaxQDA software. The categories of the content analysis were inductively derived from the interview transcripts and reviewed by both authors. The audit method (Akkerman et al., 2008) was used to ensure the quality of the data analysis. Accordingly, seven steps are required: 1.) the *orientation to the audit procedure*, in which the auditee (first author of this paper) and the auditor (second author of this paper) discuss general objectives and rules of the audit, 2.) the *orientation to the study* was not necessary as both members are part of this project, 3.) the *determination of the auditability* of the study was discussed, i.e. the extent to which the audit procedure is applicable, 4.) the *negotiation of the contract*, where the logistics were discussed, e.g. how to work together in MaxQDA and what the codes mean, 5.) the *assessment*, the auditor reviewed the coded documents and made notes of matches and discrepancies for each coding, 6.) the *renegotiation between the auditor and the auditee*, where each code was discussed and adjusted if necessary, 7.) the *auditor reported* verbally on the trustworthiness of the study.

The interview transcripts were anonymized. That means names, institutions, special subjects, etc., which were mentioned by the interviewees and could be traced back directly to a person were replaced by @@Name ## to preserve both anonymity and contextual clarity. The research data can be found in Schulz & Jacob (2025).

# Results

## Qualitative Results

During the analysis of the transcripts, ten main categories were formed (see Fig. 1), which can be derived inductively from the transcripts (see Appendix for guidelines). It was found that not all interview partners (in the following abbreviated as I1, I2 and I3) gave an answer to every question that was coded, as the answers given in this case addressed other categories in which they were coded. A total of N = 64 codings were generated from the interviews, with the individual interviewees making between 18 and 27 statements that were recorded as codings. Multiple responses from one person to the same code were possible and these were coded accordingly. In the following figures the code system with categories and subcategories are given (left side of the figure). On the right side for each interview partner (I1 to I3) a column represents their answers and how many statements they gave in this category. In Figure 1 the sum of all codings and its spread is also given as an overview. In the following figures this quantification is omitted to avoid the danger of premature quantification and generalization.

| Code System | I1 | I2 | I3 | SUM |
|---|---|---|---|---|
| Supervision of bachelor/master theses | 2 | 3 | 3 | 8 |
| Level of knowledge of lecturers regarding RDM | 4 | 3 | 3 | 10 |
| Demand for further training in RDM for students | 2 | 2 | | 4 |
| Importance of RDM for students | 8 | 10 | 2 | 20 |
| Further training in RDM by lecturers for students | 1 | 3 | 1 | 5 |
| Requests for further training regarding RDM (for students) | | | 4 | 4 |
| Requests for further training regarding RDM (for lecturers) | 1 | 1 | | 2 |
| Format requests for further training on RDM (for students) | | 1 | 3 | 4 |
| Format requests for further training on RDM (for lecturers) | 1 | 1 | 2 | 4 |
| Ideas for events for RDM inputs | | | 3 | 3 |
| SUM | 19 | 27 | 18 | 64 |

Figure 1: Code system main categories

Firstly, the lecturers were asked how they supervise Bachelor's and Master's theses in order to gain an impression of the procedure and the role of RDM. The subcategories are broken down in Fig. 2. For example, I1 explicitly described that only "individual supervision" is regularly provided. Furthermore, the subcategory that there is "no mandatory counselling" was extracted. I1 described this as follows: "That is, there are some students who then come to my consultation hours quite regularly to prepare a topic and select the topic, right up to the realization of the work, or also make separate appointments. And others pick a topic at some point and slam their work on my desk at some point and I'm just surprised that they haven't picked up any supervision in between." Two of the interviewees also described a fixed counselling scheme, which is not the same as the previous subcategory. I3 described this as follows: "Well, I roughly supervise three ideal-typical, different, three ideal-typical forms of Bachelor's and Master's theses. The first is theoretical work, the second is conceptual work and the third is empirical work, so the students who are interested in writing in my field of work are presented with this and then they decide to do one of these three forms. And now I don't know, with research data management, the empirical work is probably the most interesting. And so, of course, the supervision differs within these three types, because each requires a special form of supervision and, of course, empirical work involves a bit more organizational effort with declarations of consent and things like that. I would say that's why I think it requires more supervision than the other two forms." In addition, the provision of material for students to help them find a topic and complete their thesis is also described. Only one interviewee (I3) explicitly described the supervision of RDM content in a thesis, which was already recognizable in the last quote from I3.

| Code System | I1 | I2 | I3 |
|---|---|---|---|
| Supervision of bachelor/master theses | | | |
|   Individual supervision | | ■ | |
|   No compulsory guidance | | ■ | ■ |
|   Fixed guiding scheme | | | ■ ■ |
|   Provision of material for students | | | ■ |
|   Guidance regarding RDM | | | ■ |

Figure 2: Subcategories for the supervision of Bachelor's and Master's theses

Regarding the lecturers' level of knowledge of RDM (Fig. 3), different categories are used for this assessment. Both a low level of knowledge and a heterogeneous level of knowledge were categorized. For example, the following statement by I1 was categorized as a low level of knowledge: "That we also write into research proposals from the start that we prepare the data in the course of the project in such a way that it is also made available to others. [...] This is relatively new, I'm now gradually gaining experience myself to be honest and have to seek advice on how this works best, how to best prepare data so that it is readily available to others, how to ultimately have a data management system that not only works for your own project, but also beyond that." This quote shows that the lecturer is unsure whether RDM is implemented adequately and would like more information on this. The level of knowledge is further concretized in that there is knowledge of basic concepts and knowledge of legal aspects. I3 described this as follows: "… also the legal aspects. When you do surveys in schools. Then, of course, copyright is very relevant: what am I allowed to do with certain forms of data, and what not?". Furthermore, one interviewee described how RDM is already being used within research and explained this as follows "Well, we have always practised research data management. So you can't do empirical research without research data management. That doesn't work at all. Otherwise, my <<zeros and ones>> will end up flying across my desk. So that doesn't work. I need some kind of plan for how I collect data, how I make data accessible, how I process data, how I analyse data at the end, how I also preserve data" (I1). Thus, the (heterogeneous) knowledge of the lecturers seems to extend particularly to the area of research and data collection.

Another interesting aspect raised by I2 is that there is little experience with data publications among lecturers: "And what I think there is still little awareness of, I think overall, at least that's my feeling, is that few people do this, that you can also publish the corresponding data for specific papers that you publish somewhere, e.g. in the Open Science Forum or in other repositories. That data or the code used for statistical analyses or something like that is also stored somewhere. And to make everything as transparent as possible with the idea of open science. I would assume that there is still little experience in this area. And I don't know whether there is little knowledge, I can't judge that, but I think there is little experience of people doing it themselves in this area." According to this statement, some central aspects of RDM, such as the publication of data, are being implemented by this interviewee, which may be related to their level of knowledge.

| Code System | I1 | I2 | I3 |
|---|---|---|---|
| ∨ Level of knowledge of lecturers regarding RDM | | | |
|    Low level of knowledge | ■ | | ▪ |
|    Heterogeneouse level of knowledge | | ■ | |
|    Knowledge regarding basic terms | | | ▪ |
|    Knowledge regarding legal aspects | | | ▪ |
|    Research-internal application of RDM | ■ | | |
|    Low experience with data publication | | ▪ | |

Figure 3: Subcategory on the level of knowledge of lecturers regarding RDM

The lecturers were then asked how they assessed the students' need for further training in RDM (Fig. 4). In this context, one interviewee mentioned that the systematic teaching of RDM in the teacher training programme in so-called research workshops would be suitable, but that the research component is hardly integrated into the teacher training program: "But I know, well, we have the great difficulty with all methodological things anyway, that this is of course always a bit far away for teacher students from the actual professional goal, which is what they, what a large proportion of people who study to become teachers actually aspire to and which is already quite good, that they are all more or less forced by the research workshops to somehow do research themselves or to approach it. And in this context, this could perhaps be taught a little more systematically. There is now also a methods lecture in the new teacher training programme, but it is very small [...] (I2)". In addition, it is noted that further training in RDM would be important for final theses such as Bachelor's and Master's theses, because they are then also required: "So I think you have to teach such things when they are needed. In other words, most students, if they do any empirical research at all in their theses, do it with qualitative data (I1)".

A general need for further training in RDM in the Master's degree programme is also mentioned so that students become familiar with open data and science: "But the need, purely in terms of the idea of reusing data, and again the idea of open science, which is often linked to this, I think it would be important for students to learn something about it. But I don't really see any room for it in the current teacher training programme, except perhaps in connection with research workshops (I2)". However, it is also directly mentioned here that there is very little room for this in the teacher training programme. The need to teach RDM in graduate schools for doctoral students is endorsed without further restrictions.

| Code System | I1 | I2 | I3 |
|---|---|---|---|
| ▽ Demand for further training in RDM for students | | | |
|     Systematic teaching of RDM during studies | | ■ | |
|     Further training in RDM for final theses | ■ | | |
|     Further training in RDM for Master's program | | ■ | |
|     Demand in graduate schools for doctoral students | ■ | | |

**Figure 4: Subcategory need for further training on RDM for students**

Lecturers were then asked about the importance of RDM for them. I1 described a low significance for students in teacher training: "Because, at least I work in the teaching profession, in teacher training, I would say that, apart from the aspect of basic education, I don't see the function for my teacher student. (I1)"

I1 also stated that a "basic education in data" is important, but that the "limits of university education" must also be considered: "Yes, well, I think that part of it today is more about digital literacy, if you like. That you say, ok, every person who leaves university should actually know something about digitalisation, about data, about his or her data, about the way we deal with data today, so as a kind of [...] basic education element. But that doesn't always necessarily have to be functional for your own research. I often think that's overshooting the mark. So, it makes no sense if I structure a teacher training programme in such a way that at the end of the first state examination everyone is a little empiricist and never needs it again afterwards. (I1)". I2, on the other hand, describes the importance of this for teacher students: "So perhaps, I already said this implicitly earlier to emphasise it again. I think for many students, especially those studying to become teachers, who really have hardly any idea of how research actually works, I would say that research data management is a totally important topic, but they have to learn the basics first." I3 also describes a high level of importance for students: "If I had to name a scale, so let's go from 0-10, then I would say that this has a very high value of 9 to 10, because there are simply almost no more areas where no data is generated or where data has to be handled. And that's why I would say that the need is very high." It is also described that the importance of the Master's degree programme in particular is high. Furthermore, I2 states that it is generally difficult to weigh up relevance in the degree programme, which topic is particularly relevant and must be integrated into the curriculum, as the curriculum is already very full. I1 also mentions the fact that the "data quality in theses" is sometimes too low to process them in the sense of RDM: "And the qualitative data often does not have the substance that I would say that you really have to keep all the data from every Master's thesis or something similar. Yes, at least not so systematically. In the end, the students themselves are also responsible for this […] in my opinion. But for those students who want to stay in research, possibly do a doctorate, the situation is different. They actually have to be introduced to questions of data management and data analysis at a relatively early stage, and that's what they do." (I1) I1 makes the distinction that good data management should already be learnt in the final thesis if the students want to continue working in research.

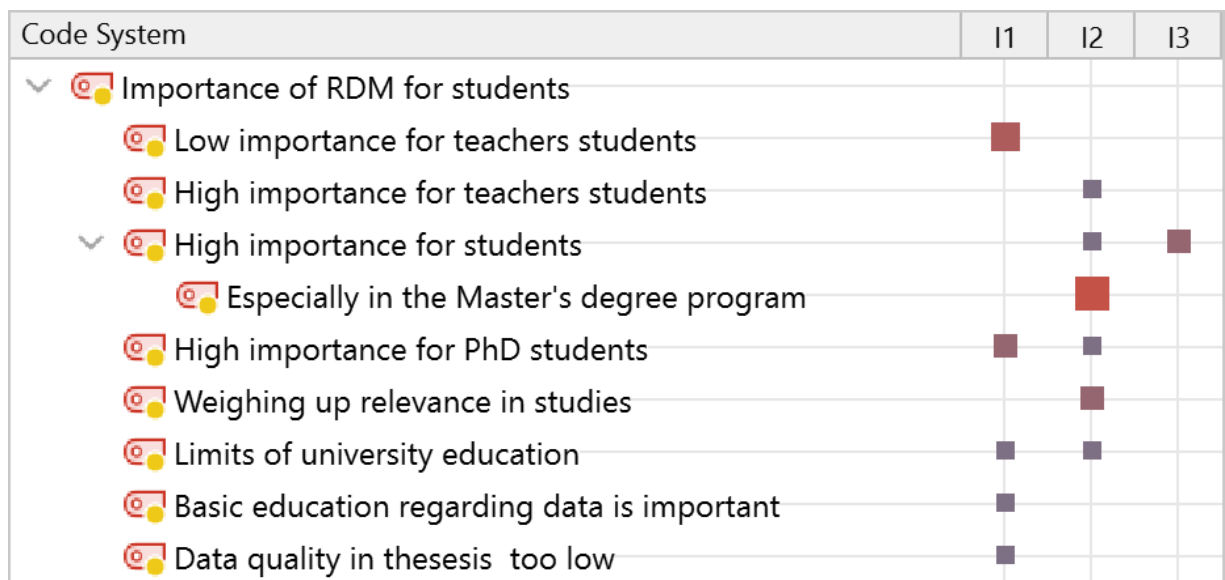

Figure 5: Subcategory importance of RDM for students

The next question asked by the interviewer was aimed at whether the lecturers impart RDM content to their students (see Fig. 6). It was found that the lecturers "do not offer in-house training or do not offer systematic teaching on RDM". I2 also describes that *RDM* is "offered as a marginal topic" and is a "potential topic in research workshops". I2 says: "Because the people in the research workshops especially the teacher students, they really have almost no methodological background at all, they are primarily concerned with other things. They first have to understand how a study actually works, what is the research methodological process? What is operationalisation? And there are completely different, much more basic things that they have to learn first and then, of course, there are always issues of transparency, open science and research data management that are somehow indirectly linked to this. "

| Code System | I1 | I2 | I3 |
|---|---|---|---|
| ▾ Further training in RDM by lecturers for students | | | |
|    No in-house training offered | ■ | | |
|    No systematic teaching in lectures regarding RDM | | ■ | |
|    RDM as a peripheral topic | | ■ | |
|    Potential topic in research workshops | | ■ | |
|    Offers from university institutions | | | ■ |

Figure 6: Subcategory: Continuing education programmes for students by lecturers

In the interview, the lecturers were also asked what further training measures they would like to see with regard to RDM. I3 described that further training should be offered in particular with "regard to the data life cycle in schools": "So everything […] that is related to this data collection cycle in the context of empirical studies in schools. I would say that's pretty important. So what types of data am I even allowed to collect in schools, how can I collect them in such a way that data protection compliance is established, i.e. that data protection is mapped." In addition, further training measures should also be created "with regard to the conscious handling of data" as well as "copyright and licences". I3 explained this in more detail with: "But in my @@specialism##, i.e. @@seminar type##, that means the digital mapping of the room. There are videos, there are pictures. Which ones am I allowed to use? Perhaps from curated platforms. There are different licences, so what do these CC licences all mean? Can I just share it like that? Can I cut something out of it? So that would be very practical things that I think would be very interesting. And I'm not a lawyer either, so I try as hard as I can, but of course I would also like to have more certainty in that respect.".

| Code System | I1 | I2 | I3 |
|---|---|---|---|
| ▾ Requests for further training regarding RDM (for students) | | | |
|    With regard to the data life cycle in schools | | | ■ |
|    With regard to the conscious handling of data | | | ■ |
|    With regard to copyright and licenses | | | ■ |

Figure 7: Subcategory wishes for further training measures on RDM for students

Lecturers were also asked which further training measures regarding RDM they would like to see for themselves (Fig. 8). I1 describes that there is "no time available for their own further training in RDM". This is explained in more detail with the statement: "I have too many tasks, which means I always have to learn such things <<on the fly>> from my people. They have to come up with the innovation, bring it into my working group and I ultimately have to gradually benefit from it." I2 is in favour of "teacher training taking priority over student training": "Because young academics should of course come into contact with such issues as early as possible and perhaps also other training opportunities, for lecturers and researchers and not so much for students first. Because if the lecturers don't know much about it, they can't teach it. And I think there's still a lot of catching up to do in many areas and yes, it's more a case of offering training programmes. How these are then accepted is of course another question, it won't be compulsory. That's the right thing to do. But somehow, there are attractive programmes for researchers and, as I said, for young researchers, especially with a view to the future in graduate school or similar contexts.".

| Code System | I1 | I2 | I3 |
|---|---|---|---|
| ∨ Requests for further training regarding RDM (for lecturers) | | | |
|     No time capacity for own further training | | ■ | |
|     Teacher training comesfirst before students | | ■ | |

Figure 8: Subcategory "Wishes for further training measures" (for lecturers)

The respondents were then explicitly asked in what format the continuing education content should be offered to students (Fig. 9). There was support for "short inputs", e.g. I2: "Yes, I think that for students, short inputs like this, if you think about the realisation, are actually quite good, yes, to raise awareness of the topic in the first place. I don't think that many students will sit down for a two-day workshop like this, maybe I'm underestimating it. But that would be the first thing, unless they are research assistants somewhere, who are perhaps motivated by the departments where they work to do so because they need it for their work there or something similar. That's why I think it would be good to start by raising awareness of the topic in a short presentation. And you could consider whether you could perhaps even incorporate it into large lectures by taking a quarter of an hour or so. And try to reach as many people as possible by giving a short input." "Short workshops" are also requested. I3 suggested that the topics should follow a didactic structure according to the data life cycle: "I think I would like it to be structured according to the data life cycle, i.e. options for data generation, data curation and storage. And that these different possibilities are then shown within this. So how can I obtain data in the school context? Or what data is generated anyway? Or how can teachers also obtain such data? All the way to: How do I store it, where do I store it? How can I visualise it? How can I interpret it? Yes, I would think along these lines."

| Code System | I1 | I2 | I3 |
|---|---|---|---|
| ∨ Format requests for further training on RDM (for students) | | | |
|     Brief inputs | | ■ | ■ |
|     Short workshops | | | ■ |
|     Didactic structuring according to the data life cycle | | | ■ |

Figure 9: Format requests for continuing education measures RDM (for students)

In order to concretise a possible offer, the lecturers were asked which formats they would like to see with regard to continuing education measures in RDM (Fig. 10). Lectures in research colloquia for doctoral students and programmes in graduate schools were mentioned. Beyond the target group of doctoral students, workshops for lecturers and templates for lecturers/researchers that they can simply use were also mentioned: "What we haven't really discussed is this data management plan, because it has of course been hugely relevant in research recently or for some time now. In some cases, you can no longer apply for projects if you don't submit it, especially for EU funding. You also have to make this clear to the DFG. That's why I would like to reiterate this, also on the part of the lecturers. Of course, this doesn't affect university teaching in that sense, but rather university research. But I would also think it would be good if there were workshops on this. With sample [...] data management plans. If you look at it that way: How do successful applicants organise this? Not in terms of teaching, but I just wanted to let you know. I've just realised that again" (I3).

| Code System | I1 | I2 | I3 |
|---|---|---|---|
| Format requests for further training on RDM (for lecturers) | | | |
|    Presentations at doctoral colloquium for PhD students | ■ | | |
|    Offers in graduate schools | | ■ | |
|    Workshops for lecturers | | | ■ |
|    Templates for lecturers/researchers | | | ■ |

Figure 10: Format sub-scale for further training measures RDM (for lecturers)

The previously mentioned measure of RDM inputs (Fig. 11) was discussed in more detail with I2, who was asked about suitable events. I2 suggested that the introductory course in educational science could be suitable for this: "Well, the big introductory lecture in educational science, the one that, as far as I know, all educational scientists and all teacher students have to go through. And in principle, that would have reached everyone in theory." A separate RDM lecture was also proposed, as well as RDM as a component of research workshops. The research workshops take place as an elective course, although I2 suggested that all students should still be reached: "And what perhaps makes sense: In the Master's programme in the teacher training programme, if you just add that in the context of research workshops. Because that's where the students in the teacher training programme really must deal with research in concrete terms. And of course there are many, many parallel courses, I don't even know how many there are, 30 or so. I don't know now, you certainly can't afford to go to all of them. But perhaps we can also find a format at some point to organise a joint event, perhaps online or something. Because at least they're all in the same time slot, the research workshops, so maybe that could even be implemented. Exactly, and to be honest, I currently have very little to do with the Master's programme in educational science."

| Code System | I1 | I2 | I3 |
|---|---|---|---|
| Ideas for events for RDM inputs | | | |
|    Introductory lecture in the educational sciences | | ■ | |
|    RDM lecture | | ■ | |
|    Part in research workshops | | ■ | |

Figure 11: Subscale event proposals for RDM inputs

## Quantitative Results

Figure 12 shows the responses of three interviewees to questions on the relevance of aspects of research data management for students and lecturers on teacher training programmes. The left-hand column shows the aspects that were asked about. The right-hand column shows the respondents' answers on the relevance of RDM topics for students and lecturers. The data was collected in the same interviews described in the qualitative data analyses, but in this part the lecturers got a table with RDM topics and were requested to tick the relevance for students and lecturers of the single topics. The answers of the interview partners are summarized in Figure 12. The blue bars encompass the relevance of students and the green bars the relevance for lecturers. The scale 0 to 3 shows how many interviewed lecturers rated the specific RDA topic as relevant ("Yes, this is relevant for this group").

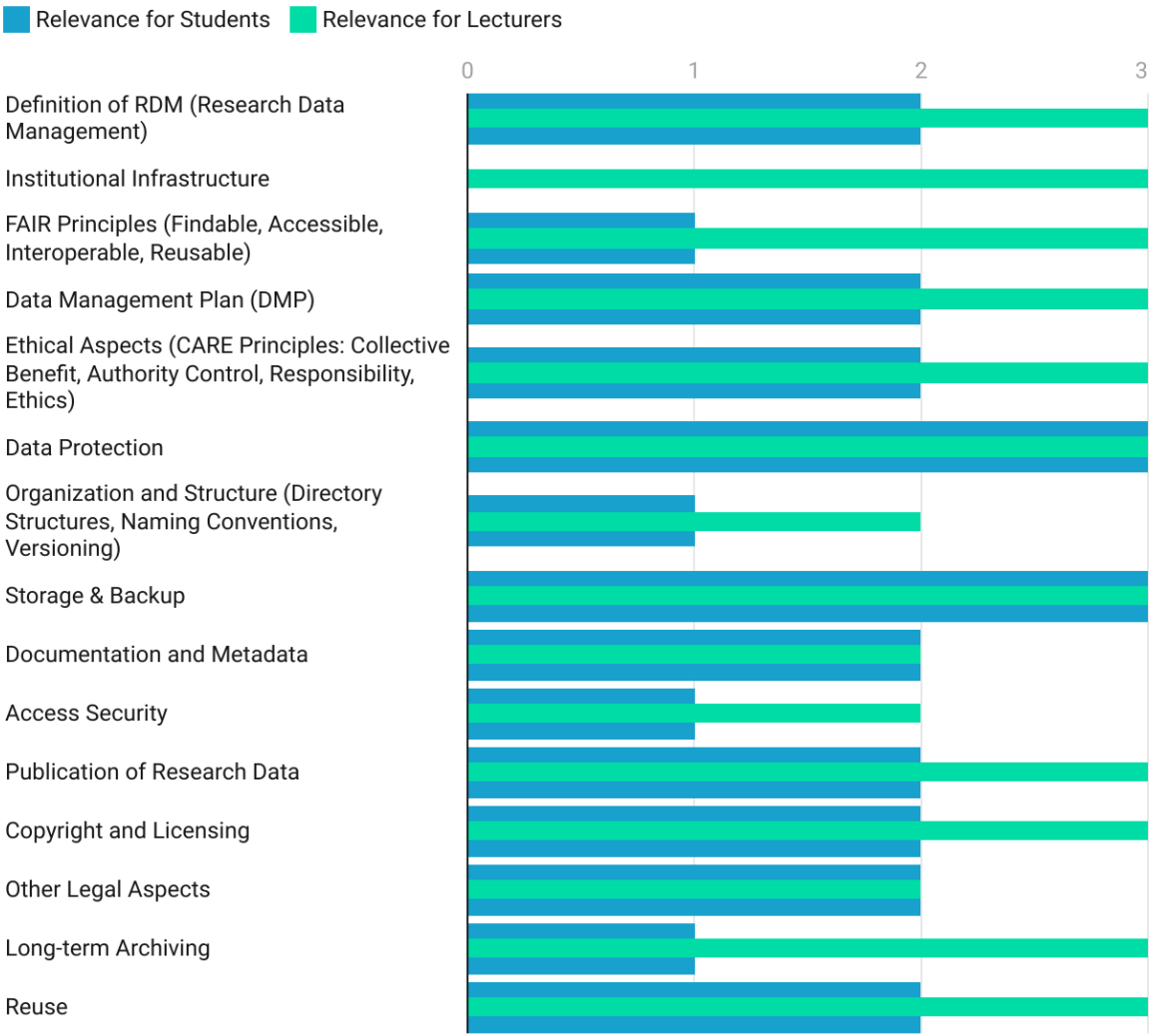

Figure 12 Relevance of RDM topics rated by lecturers

For the students, the interviewees see the aspects mentioned as less relevant overall and several points are only stated as relevant by one person, including institutional infrastructure, FAIR principles, organisation and structure and access security. There is therefore a clear overlap with the points that are less relevant for lecturers.

To illustrate the different perspectives of the interviewees, Figure 13 presents each individual's responses to the various RDM topics. When considering the relevance of RDM for students, it is noticeable that I1 only considers a few aspects to be relevant (e.g., data protection, other legal aspects). In contrast, I2 and I3 evaluate almost all the mentioned categories as relevant. However, I2 excludes the topics of organization and structure, access security, and other legal aspects. I3 only excludes institutional infrastructure and the FAIR principles. On the educators' side, almost all aspects are regarded as relevant. Only a few topics are excluded, such as organization and structure, documentation and metadata, access security, and other legal aspects. This is where there is a stronger consensus among the interviewees.

## Rated Relevance of RDM Topics per Interviewed Person

Relevance for Students (I1) | Relevance for Students (I2) | Relevance for Students (I3) | Relevance for Lecturers (I1) | Relevance for Lecturers (I2) | Relevance for Lecturers (I3)

| | Relevance for Students (I1) | Relevance for Students (I2) | Relevance for Students (I3) | Relevance for Lecturers (I1) | Relevance for Lecturers (I2) | Relevance for Lecturers (I3) |
| --- | --- | --- | --- | --- | --- | --- |
| Definition of RDM (Research Data Management) | 0 | 1 | 1 | 1 | 1 | 1 |
| Institutional Infrastructure | 0 | 0 | 0 | 1 | 1 | 1 |
| FAIR Principles (Findable, Accessible, Interoperable, Reusable) | 0 | 1 | 0 | 1 | 1 | 1 |
| Data Management Plan (DMP) | 0 | 1 | 1 | 1 | 1 | 1 |
| Ethical Aspects (CARE Principles: Collective Benefit, Authority Control, Responsibility, Ethics) | 0 | 1 | 1 | 1 | 1 | 1 |
| Data Protection | 1 | 1 | 1 | 1 | 1 | 1 |
| Organization and Structure (Directory Structures, Naming Conventions, Versioning) | 0 | 0 | 1 | 1 | 0 | 1 |
| Storage & Backup | 1 | 1 | 1 | 1 | 1 | 1 |
| Documentation and Metadata | 0 | 1 | 1 | 1 | 1 | 0 |
| Access Security | 0 | 0 | 1 | 1 | 0 | 1 |
| Publication of Research Data | 0 | 1 | 1 | 1 | 1 | 1 |
| Copyright and Licensing | 0 | 1 | 1 | 1 | 1 | 1 |
| Other Legal Aspects | 1 | 0 | 1 | 1 | 0 | 1 |
| Long-term Archiving | 0 | 0 | 1 | 1 | 1 | 1 |
| Reuse | 0 | 1 | 1 | 1 | 1 | 1 |

Created with Datawrapper

**Figure 13 Relevance of RDM topics rated by lecturers per person**

# Discussion and Conclusion

When analysing the **qualitative data**, it is noticeable that the lecturers rate their own level of knowledge of RDM as low. In particular, they have little experience with publishing data. This may be related to the fact that publishing research data does not yet receive significant scientific recognition (Kiesler & Schiffner, 2022). Nevertheless, the lecturers report that they need this knowledge for third-party funded projects and their applications and that they deal with it, albeit not very systematically. Notably, they do not feel sufficiently confident in their RDM skills to effectively teach these concepts to students. When it comes to RDM instruction, they tend to refer students to central university services rather than providing the content themselves. Related research also shows that lecturers often have difficulties categorizing research data and need more guidance to support open science practices (Kiesler et al., 2024).

Overall, the lecturers in the interviews rate the importance of RDM for students as low, although its relevance for Master's degree courses is increasing. This is certainly due to the fact that the Master's course involves significantly more work with research data through so-called research workshops and an often empirical Master's thesis, or that students collect their own research data. When specifying further education requirements for students, some list rather abstract categories such as handling data, but more specific topics such as copyright and licenses are also mentioned.

Regarding the format requirements for lecturers and students, it is noticeable in the interviews that there should be courses with different durations and for different levels. Possible content for this can be taken from the existing learning objectives matrix (Petersen et al., 2023). The lecturers also emphasize that templates should exist that make it easier to deal with the topic and provide advice on it. In principle, however, the lecturers point to the problem that most students' timetables are already full and that there is a lot of competing content that should be taught on a compulsory basis. They agree that students should receive credits for taking RDM courses, but they also acknowledge that it is difficult to make this a mandatory part of their studies. Free elective courses offer a good opportunity in the teacher training programme, but not all students can be reached with this approach. In addition, lecturers are free in their research and teaching and cannot be obliged to provide further training in RDM themselves. However, it is certainly possible to make their work easier by providing good handouts and adapting and distributing them in the relevant communities. In addition, further training courses for lecturers on RDM in their research could raise their awareness of students' need for RDM skills.

However, as described by Qin and D'Ignazio (2010), "Lessons learned from a two-year experience in science data literacy education", show that it is important to weigh the merits of a "pull" strategy (e.g. experts presenting data management in workshops) against those of a "push" strategy (e.g. flyers posted in campus buildings or short presentations and Q&A sessions as guests).

The analysis of the **quantitative data** shows a very different assessment of relevance for students. For example, the interviewees excluded topics such as the institutional infrastructure as relevant for students, which is, however, very important for storing data. If students use the university services, it is essential that they know, for example, how often backups are created and how securely their data is protected against loss.

In contrast, the relevance of the different topics is clearer for teaching staff, with interviewees estimating that almost all topics are important for this group. However, topics such as documentation and metadata, access rights, organization and structure as well as other legal aspects were considered less relevant by the interviewees. Yet, these topics are essential for the successful implementation of large projects in which a lot of (sensitive) data is generated.

When **comparing the qualitative and quantitative data**, it is noticeable that the lecturers should know about all topics in RDM (see quantitative analysis), but the qualitative part shows that they feel rather insecure and sometimes lack the time to engage with these topics in depth. It is particularly striking that one of the lecturers (interviewee I1) considers the relevance of RDM for students to be rather low when asked about the term RDM. However, as soon as they are confronted with the specific contents of RDM, they seem to rate its relevance higher. Of course, it cannot be expected that lecturers are experts in RDM. Nonetheless, the interviews suggest that those who are already knowledgeable about RDM tend to perceive its relevance for students as particularly high.

In order to enhance RDM competencies, it is inevitable to educate lecturers and students. Some formats were discussed in this paper.

This work is limited by the following factors. As is usual in qualitative research, only a few respondents were interviewed. It is not the aim here to make generalized statements. Nevertheless, further interviews should be conducted, and other subjects and locations should be included to obtain as many different perspectives as possible and to be able to sketch a comprehensive picture of the situation at universities. A comparison with other countries should also be made in the future.

Furthermore, the interviewees were given various RDM categories in the interview on which they positioned themselves. It cannot be ruled out that the naming of the categories had an influence on the interviewees, nor that the interviewees had too little knowledge of RDM themselves to be fully aware of the topics included within the categories. As there are also different matrices for categorization, this work depends on the matrix chosen. However, this exploratory work provides initial empirically based insights into challenges and potentials to integrate DL and RDM into study curricula.

Recommendations for actions in teaching and curriculum design:

- Lecturers should work together closely with their local research data management centre to foster their own RDM competencies
- Handouts for lecturers and students need to be provided to help lecturers advise students regarding RDM topics (e.g. regarding informed consent when gathering research data, anonymization of data, data management plan, etc.)
- Compulsory DL or RDM courses are hard to integrate in already busy curricula. Spiral curricular approaches within the subject being studied appear to be easier to integrate, as shown in (Biernacka & Schulz, 2022), for example.

# Appendix

**Interview guide for lecturers**

**Objectives**

- to determine the need for RDM education in the relevant subject areas
- to determine the level of knowledge of RDM among lecturers (department-specific?)
- to determine the importance of RDM for lecturers (How important do lecturers consider RDM or the related training of teacher students in their department to be in this regard?)
- determination of the (further) training measures taken by lecturers (for themselves and their teacher students)

**Interview questions**

1. How long have you been teaching at the University of Hamburg?

2. Please describe the process of supervising bachelor and master theses in your department

3. Please describe the level of knowledge of lecturers regarding RDM in your department

4. Please describe the need you see for training measures for teacher students in your department regarding RDM.

5. Please describe the importance of training for teacher students regarding RDM.

6. Please explain what the lecturers in your department are doing specifically for the RDM education of teacher students.

7. Please describe the educational activities you would like to see from the CRDM. (With regard to lecturers and teacher students)

8. Can you describe any specific wishes regarding the didactic implementation?

    a. Short input (presentation of CRDM) approx. 15 min.

    b. Lecture/ seminar (1.5 hours)

    c. Workshop (3 hours to 2 days)

9. Do you have specific events/ series of events or other frameworks for such CRDM programmes?

*I have brought a list of RDM topics with me and would like to know your views on the following two points*

a.   Relevance for teacher students in your department in relation to the following topics

b.   Relevance for lecturers (or their research) in relation to the following topics

List of topics

    1. definition of RDM (research data lifecycle)

    2. institutional infrastructure (support for RDM, tools)

    3. FAIR principles (Findable, Accessible, Interoperable, Reusable)

    4. data management plan (contents of the DMP)

    5. ethical aspects (CARE principles - Collective Benefit, Authority Control, Responsibility, Ethics)

    6. data protection (personal data, informed consent)

    7. organisation and structure (directory structures, naming convention, versioning)

    8. storage & backup

    9. documentation and metadata (content and forms of documentation, standards)

    10. access security (encryption, password protection, rights management)

    11. publication of research data (data selection, publication options)

    12. copyright and licensing

    13. other legal aspects (patent law, contractual agreements)

    14. long-term archiving (definition of the term, sustainable file formats, requirements for long-term archiving)

    15. reuse (finding research data, terms of use, access rights)